\def\<{\langle}
\def\>{\rangle}

\documentclass[pra,aps,twocolumn,tightenlines,showpacs]{revtex4}
\usepackage{epsfig}

\begin{document}
\title{The Bell Inequality: A measure of Entanglement?}
\author{W. J. Munro, K. Nemoto and A. G. White}
\affiliation{Special Research Centre for Quantum Computer Technology,
The University of Queensland,  Brisbane, Australia}
\date{\today}
\begin{abstract}
Entanglement is a critical resource used in many current quantum information
schemes.  As such entanglement has been extensively studied in two qubit systems
and its entanglement nature has been exhibited by violations of the Bell inequality.
Can the amount of violation of the Bell inequality be used to quantify the
degree of entanglement. What do Bell inequalities indicate about the nature
of entanglement?
\end{abstract}
\maketitle

Entanglement was recognised early as one of the key features of quantum
mechanics\cite{EPR 35,Schrodinger 35}. Entanglement can be described
as the correlation between distinct subsystems which cannot be
created by local actions on each subsystem separately. The
advantage offered by quantum entanglement relies on
the crucial premise that it not be reproduced by any
classical theory\cite{Bell 65,CHSH 69,Clauser and Shimony 78}.  Despite the fact that the
possibility of quantum entanglement was acknowledged almost as
soon as quantum theory was discovered, it is only in recent years that
consideration has been given to finding  methods to  quantify
it\cite{Bennett 96,Bennett 96a,Bennett 96b,Peres 96,Popescu 97,Horodecki
96,Horodecki 97,Hill 1997,Vedral 98,Wootters 1998,Rungta 2000,Deuar 2000}.
Historically the Bell inequalities were seen as a means of
determining whether a two qubit system is entangled.  It was
known that the larger the violation of the Bell inequality the more
entanglement present in the system\cite{bellmeasure}. This lead to the
perception that to some degree the Bell inequalities were a
measure of entanglement in such systems.

In 1994 it was discovered that not all entangled states violate a Bell
inequality\cite{Popescu 94}. It was shown that the Werner state, a mixture
of the maximally entangled state and the maximally mixed state can be
entangled (inseparable) and yet not violate the conventional Bell
inequality\cite{Werner 89,Horodecki 95}. It was found that multiple
copies\cite{Peres 96} of the Werner state
could be distilled to a state that does violate the Bell inequality.
Hence it is important to specify now that our interest lies in
whether a single particular state violates such an inequality.
It was shown that by Gisin {\it et. al}\cite{Gisin 96}
that there are states which do not violate this inequality,
but can be distilled by local operations and classical communication
to produce a state that does. Our interest is in whether the original
state violates such inequalities. These observations are
important experimentally because the Bell
inequalities have been one of the few methods available to determine
whether a two qubit state has quantum properties. There have been quite a
number of experimental tests of the Bell inequality\cite{general Bell}
using  polarisation entangled photons from a spontaneous parametric
down conversion source. Kwiat {\it et al.}\cite{Kwiat 99} have
shown a $242\sigma$ violation of Bell's inequality. Here maximally
entangled pure Bell states were produced, however these sources are
currently being used to synthesise two qubit polarisation quantum
states, with a variable degree of entanglement and purity\cite{White 2000}.
The question then becomes how do we characterise the entanglement
in such systems. In the current work such states are being characterised
by quantum state tomography which allows the reconstruction
of the reduced density matrix for the polarization entangled
photons\cite{White99,White 2000}.  Hence one can determine all
the physically relevance properties such as the degree of entanglement
and purity. The process to reconstruct the density matrices requires
many more measurements than those required to violate a Bell inequality.

We will structure this article as follows. We will begin by defining
how we will characterise our two qubit state in terms of its degree
of entanglement and degree of mixture. We will then proceed to specify
the particular Bell inequality we will consider here. At this point here we
comment again that we are interested in tests that can be performed on
a single entangled pair of qubit (primarily because this is physically
currently). There are a number of possible measurements that can be
performed but here we will restrict our attention to POVM's. We will not
consider generalised measurements. Given these tools we now examine
the degree of violation versus the degree of entanglement
for two classes of states; the Werner state\cite{Werner 89} and
the maximally entangled mixed state\cite{Munro 2000}. We will attempt to answer
the question as to ``what the Bell inequality
indicate about the nature of entanglement?''. Is it only weakly
entangled states that do not violate such inequalities?

In examining the degree of entanglement there are currently a
number of measures available. These include the entanglement of
distillation\cite{Bennett 96}, the relative entropy of
entanglement\cite{Vedral 97}, but the canonical measure of
entanglement is called the entanglement of formation
(EOF)\cite{Bennett 96} and for a pure state is simply given by the
von Neumann entropy\cite{vonNeumann55} of that reduced density
matrix. For a mixed state $\hat \rho$ the entanglement of
formation is defined as,
\begin{eqnarray}
E_F(\hat \rho)=\min \sum_i p_i E_F(\psi),
\end{eqnarray}
where this minimum is taken over all the possible
decompositions of $\hat \rho$ into the pure states
$\hat \rho=\sum_i p_i|\psi_i\rangle\langle\psi_i|$. The entanglement
of formation for an arbitrary two qubit system has been found by
Wootters\cite{Hill 1997} to be simply given by,
\begin{eqnarray}\label{EOF}
E_F(\hat \rho)=h\left(\frac{1+\sqrt{1-\tau}}{2}\right),
\end{eqnarray}
where $h(x)$ is Shannon's entropy function,
\begin{eqnarray}
h(x) = -x\log(x) - (1 - x)\log(1 - x),
\end{eqnarray}
and $\tau$ is the tangle\cite{Coffman 99} (concurrence
\cite{Wootters 1998} squared). The tangle $\tau$ is given by,
\begin{eqnarray}
\tau={\cal C}^{2}=\left[\max\{\lambda_1-\lambda_2-\lambda_3-\lambda_4,0\}\right]^{2}.
\end{eqnarray}
where the $\lambda$'s are the square root of the eigenvalues in decreasing
order of,
\begin{eqnarray}
\hat \rho \tilde \rho = \hat \rho\;\sigma_y^A \otimes \sigma_y^B
\hat \rho^* \sigma_y^A \otimes \sigma_y^B.
\end{eqnarray}
Here $\hat \rho^*$ denotes the complex conjugation of $\hat \rho$ in
the computational basis $\{|00\>, |01\>,|10\>,|11\>\}$.
With the entanglement of formation $E_F$ being a strictly monotonic
function of $\tau$, the maximum of $\tau$ corresponds to the
maximum of $E_F$. Hence the tangle may also be considered a direct measure of
the degree of entanglement and this is what we will consider in this
article. In general the measure of entanglement to be used depends
heavily on your the desired us of that information. The entanglement
of distillation may be a much more useful practical measure but it is
different to calculate in practice. In general the  entanglement of
distillation is smaller than  entanglement of formation.

For a general two qubit density matrix the purity of the state
provides complementary information about the state. The purity measure
described here is the linearised entropy\cite{Bose00} of $\hat \rho $ given
by
\begin{eqnarray}
S_{L} = {4 \over 3} \left\{1-{\rm Tr}\left[\rho^{2} \right] \right\}
\end{eqnarray}
The $4/3$ normalisation\cite{White 2000} for $S_{L}$ ensures
that for a general two
qubit density matrix $S_{L}$ ranges between $0$ and $1$. The von Neumann
entropy\cite{vonNeumann55} of the state could be used but $S_{L}$ is
easier to calculate and provides the same degree of characterisation.
With an explicit expression for the degree of entanglement and the
degree of mixture let us now turn our attention
to the Bell inequality and what a violation of it potentially indicates
about the nature of entanglement for the two qubit system.
There are a very number of Bell inequalities that could be investigated
in this article but we will focus our attention on the
{\it original} two qubit Bell inequality\cite{Bell 65,CHSH 69,Clauser and Shimony 78},
\begin{eqnarray}\label{spin}
{\bf B}_{\rm{S}}&=&\left|E\left(\phi_{1},\phi_{2}\right)+
E\left(\phi_{1}\phi_{2}'\right)\right. \nonumber \\
&\;&\;\;\;\;\;\;\left.+E\left(\phi_{1}',\phi_{2}\right)+
E\left(\phi_{1}',\phi_{2}'\right)\right|\leq 2,
\end{eqnarray}
where the correlation function $E\left(\phi_{1},\phi_{2}\right)$ is given by,
\begin{eqnarray}\label{correlation}
E\left(\phi_{1},\phi_{2}\right)&=&{\rm Tr} \left\{ \hat S_{1}(\phi_{1}) \hat
S_{2}(\phi_{2}) \hat \rho \right\},
\end{eqnarray}
with,
\begin{eqnarray}\label{correlationfn}
\hat S_{i}(\phi_{i}) &=& \cos \phi_{i} \left[|0\>\<0|-|1\>\<1|
\right]\nonumber \\
&\;&\;\;\;\;\;+ \sin \phi_{i} \left[e^{i \bar
\phi_{i}}|0\>\<1|+e^{-i \bar \phi_{i}}|1\>\<0| \right].
\end{eqnarray}

The inequality (\ref{spin}) is violated if ${\bf B}_{\rm{S}}>2$.
In the above expression the $\phi_{i}$'s are the analyser settings for
the $i^{th}$ particle ($i=1,2$). In calculating whether the Bell inequality
is violated, the choice of analyser settings is absolutely critical.
In this article we will choose them to maximise the violation for
the actual state under consideration.

It is now time to turn attention to the class of states we will
consider in this article. The Hilbert space in which two qubit reside is
large and hence in this article we will generally restrict our attention to two
particular types of states. The first state is of the form,
\begin{eqnarray}\label{wernerlike}
\hat \rho(\gamma,\xi) = {{1-\gamma}\over{4}} I_{2}\otimes I_{2} + \gamma |\Psi_{\rm non}\>\<
\Psi_{\rm non}|,
\end{eqnarray}
where,
\begin{eqnarray}
|\Psi_{\rm non}\>=\cos \xi |0\>|0\>+ e^{i \phi} \sin \xi|1\>|1\>.
\end{eqnarray}
For $\xi=\pi/4$ eqn (\ref{wernerlike}) is of the
form normally attributed to the usual Werner state\cite{Werner 89}
which was the first state found to be entangled for certain $\gamma$
and yet not violate a Bell inequality for single states.

We will refer to the state (\ref{wernerlike}) as a non-maximal Werner state
as it is a mixture of a non maximally entangled state and the fully
mixed state. In the limit of $\gamma=1$ eqn (\ref{wernerlike}) represents a non
maximally entangled pure state.  It is straight forward to show that
the mixture of the nonmaximally entangled state and the fully mixed
state given by (\ref{wernerlike}) is entangled only when,
\begin{eqnarray}
\gamma > {{1}\over{1+2 |\sin \left(2\xi\right)|}}.
\end{eqnarray}
(for the Werner state it is entangled for $\gamma >{1\over 3}$
\cite{Bennett 96b,Braunsteinetal 98}).
It is also possible to derive an explicit expression for the degree of
entanglement for such states using the tangle measure.
While it is quite complicated one can
show that the tangle for the state (\ref{wernerlike}) is given by,
\begin{eqnarray}
\tau=\left[\max\{ {{\sqrt{\Lambda_{1}}-\sqrt{\Lambda_{2}} }\over 4}
-{{1-\gamma}\over{2}},0 \}\right]^{2},
\end{eqnarray}
where,
\begin{eqnarray}
\Lambda_{1\atop 2}&=&\pm 4\gamma  \sin \left(2\xi\right)
\sqrt{{\left( 1 + \gamma  \right)}^2-4 \gamma^{2} \cos^{2} (2\xi)} \nonumber \\
&\;&\;\;\;\;\;\;\;\;\;\;+{\left( 1 + \gamma  \right) }^2-4\,\gamma^2\,\cos (4\xi).
\end{eqnarray}

The second state we will consider is the maximally entangled mixed
states recently predicted by White {\it et. al}\cite{White 2000}.
This states has the explicit form,

\begin{eqnarray}\label{mems}
\hat{\rho_{mems}}=\left(
\begin{array}{cccc}
g(\gamma)& 0& 0& {\gamma\over 2}\\
0 & 1-2g(\gamma) & 0 &  0 \\
0 & 0 & 0 & 0 \\
{\gamma\over 2} &  0  & 0  & g(\gamma)\\
\end{array}\right),
\end{eqnarray}
where,
\begin{eqnarray}
g(\gamma)=\left\{
\begin{array}{cccc} \gamma/2 & & &\gamma \geq 2/3\\ 1/3 & & &\gamma < 2/3 \\
\end{array} \right. ,
\end{eqnarray}
and has been shown to have the maximal amount of entanglement for a certain degree of
mixture (as measured by linear entropy) or vice versa.  This state is
entangled for all nonzero $\gamma$ and in fact it has been shown that
the tangle simply given by
\begin{eqnarray}
\tau= \gamma^{2}
\end{eqnarray}
For a given degree of mixture, the maximally entangled mixed state is generally
significantly more entangled that the Werner state at the same degree
of mixture.

\begin{figure}[!ht]
\center{ \epsfig{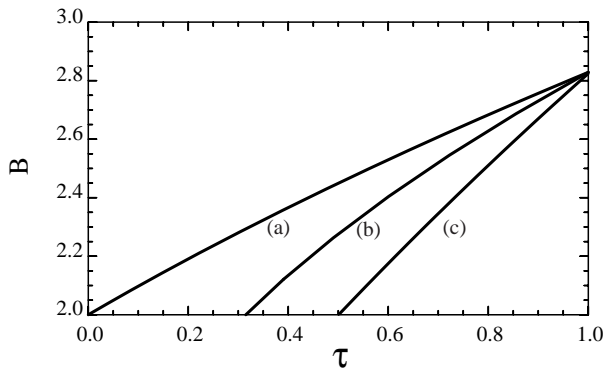}}
\caption{Plot of the maximum violation of the spin Bell inequality
versus the degree of entanglement (tangle $\tau$) for the non-maximally entangled pure
state represented by the density operator $\hat \rho(1,\xi)$ (curve a)
and the Werner state represented by the density operator $\hat
\rho(\gamma,\pi/4)$ (curve b). The analyser settings have been chosen to
maximise the violation for the particular $\gamma, \xi$ values.
A violation of the spin Bell inequality is achieved when $|B_{\rm S}|>2$.
Also shown in the figure are the results for the maximally entangled mixed state (curve c).}
\label{fig1}
\end{figure}

Let us now examine how well these two state violate a Bell
inequality. The state (\ref{wernerlike}) violates the Bell inequality (\ref{spin})
for quite a wide range of $\gamma,\;\xi$ values. In Fig. (\ref{fig1})
we plot the maximum value of ${\bf B}_{\rm{S}}$ (optimising the analyser
settings to maximise the violation) versus the degree of entanglement
(as measured by the tangle). We have plotted two specific parameter sets
\begin{itemize}
\item the nonmaximally entangled pure state $\hat \rho(1,\xi)$
\item and the Werner state given $\hat \rho(\gamma,\pi/4)$.
\end{itemize}
This results show very clearly that the Werner state (given by the
density matrix $\hat \rho(\gamma,\pi/4)$) and the non-maximally entangled pure state
(given by the density matrix $\hat \rho(1,\xi)$) violate the Bell inequality by
difference amounts for the same degree of entanglement.
In fact for these two different classes of entangled states,
there is a clear region where one of the states (the non-maximally entangled
pure state) violates the Bell inequality but the Werner state does
not\cite{Popescu 94}. This was a surprising result when it was first
found by Popescu\cite{Popescu 94}.

It showed that non all entangled states violate a Bell inequality. All pure
two qubit entangled states do violate a Bell inequality\cite{Gisin 91}
and in fact the degree of violation is equal to $2
\sqrt{1+\tau^{2}}$\cite{Sanders 00}. However as a state
becomes mixed it is harder to violate the Bell
inequality. The Werner state does not violate our Bell inequality if it
tangle is less than $\tau \leq 1/3$ ($EOF=0.44229$). This is quite a small degree of
entanglement and has lead to the perception that it is only certain
weakly entangled mixed states that do not violate the two qubit Bell
inequality. To investigate this point further less us consider the
maximally entangled mixed state that we described in (\ref{mems}).
This state has it degree of entanglement maximised for a given purity
and vice versa.

In Fig.  (\ref{fig1}), we have also plotted (curve c)
the degree of violation of the Bell
inequality versus the tangle for this maximally entangled mixed state.
We observe in this Figure that significantly more entanglement
is required to violate the Bell inequality  to the same degree for the
maximally entangled mixed state than for the Werner state. In fact our
Bell inequality for the maximally entangled mixed
state is only violated if $\tau > 0.5$
($EOF=0.6$). This is a significant degree of entanglement given
that a Bell state has $\tau = 1.0$ ($EOF=1.0$) and a separable state
has $\tau = 0.0$ ($EOF=0.0$). The maximally entangled mixed
state we have considered here
has a maximal degree of entanglement for a given linear
entropy (the choice of degree of mixture in this case). There are
other choices for the degree of mixture, not based on purity, and
these may have a tangle value $\tau > 0.5$ while still not violating
Bell inequality. This is left for further investigation.

The result above also tentatively indicate
that the more mixture contained in a state, the higher the degree of entanglement required
for it to violate the two qubit Bell inequality. What these results
indicate that if a state has a certain degree of entanglement (this
may be large), it is not possible to infer whether that state will violate
the Bell inequality or by how much. This is we believe the first instance where it
has been explicitly demonstrated via quantifiable measures that the
size of the violation of the Bell inequality for an unknown two qubit
state is not absolutely related to the degree of entanglement in that
state.

Let us now investigate in some detail the effect of mixture of our entanglement and
Bell inequality question. Again we will examine two specific states, the first
being our non-maximally entangled Werner state. We choose this state
as it has the
property that with the two parameters $\gamma,\xi$ we can change the
state from a non-maximally entangled pure state to the Werner state.
We know that the non-maximally entangled pure state $\gamma=1$
violates the Bell inequality as soon as the state contains
entanglement ($\xi\neq 0$)\cite{Gisin 91}. However the Werner state (with $\xi=\pi/4$
only violates the Bell inequality when $\tau>{1\over 3}$. We will
investigate what occurs between these two extremes.
The second state we will examine is a modification of the
maximally entangled mixed state,
\begin{eqnarray}\label{memslike}
\hat \rho_{m}(\gamma,\xi) = (1-\gamma) |0\>|1\>\<0|\<1| + \gamma |\Psi_{\rm non}\>\<
\Psi_{\rm non}|.
\end{eqnarray}
and is a mixture of the non-maximally entangled pure state and the
diagonal density matrix element $|0\>|1\>\<0|\<1|$.
As for the first state mentioned the two parameters in this state
also make it possible to change its behaviour for a nonmaximally
entangled pure state to the maximally entangled mixed state.
Choosing the parameters $\gamma$ and $\xi$ such that
both states (\ref{wernerlike}) and (\ref{memslike}) are
a non-maximally entangled pure state that just satisfies the Bell
inequality (that is ${\bf B}_{\rm{S}}=2$) we vary the parameters $\gamma,\xi$ such that we
increase the degree of mixture in the system while maintaining ${\bf
B}_{\rm{S}}=2$. For these new $\gamma$ and $\xi$ values we then determine
the degree of entanglement and mixture ensuring that ${\bf B}_{\rm{S}}=2$.
\begin{figure}[ht]
\center{ \epsfig{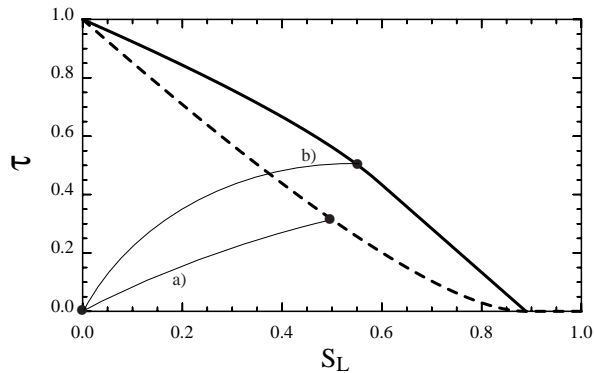}}
\caption{Plot of the degree of entanglement versus linear entropy for
the states (\ref{wernerlike}) and (\ref{mems}). The Werner state
is displayed as a dotted line while the MEMS state (\ref{mems})
is displaced as the solid dark curve. The tangle $\tau$ of the Werner
and MEMS state increases as the system becomes less mixed.
The non-maximally entangled pure state may be
represented by a line along the y-axis at a linear entropy of $S_{L}=0$.
The  non-maximally entangled pure state satisfies the Bell
inequality (${\bf B}_{\rm S}=2$) at $\tau=0$. Curve a) traces out the
curve for the state (\ref{wernerlike}) where $\gamma$ and $\xi$ are
chosen such that ${\bf B}_{\rm S}=2$. Curve b) traces out the
curve for the state (\ref{memslike}) where $\gamma$ and $\xi$ are
chosen such that ${\bf B}_{\rm S}=2$. In all situations here
the analysers setting were chosen to maximise the potential violation.}
\label{fig2}
\end{figure}
In Figure (\ref{fig2}) we plot on the tangle-linear entropy plane, the
boundary curve where ${\bf B}_{\rm{S}}=2$ for both states.
The tangle axis (y-axis) represents the degree of entanglement
while the x-axis displays the degree of mixture.  Figure
(\ref{fig2}) confirms for these states our idea that as the state
becomes more mixed, more entanglement is required to violate the Bell
inequality. If we again examine the state (\ref{wernerlike}) then
points for this state that fall below the curve a) in Figure
(\ref{fig2}) are entangled (if $\tau>0$) but do not violate the
inequality we have considered.

To summarise, in this article we have investigated the extent to
which the Bell inequality may be considered a measure of entanglement.
Our results indicate that the more mixed a system is made
the more entanglement is generally required to violate
the original Bell inequality to the same degree. We have specifically
illustrated an example where a state (the maximally entangled mixed state)
has a tangle of $\tau=0.5$ represents a significant degree of entanglement
(an $EOF=0.6$) yet does not violate the Bell inequality considered
here. This dispels the impression that it is only the weakly
entangled states that do not violate the Bell inequality.
For a specific class of state, for instance the Werner state,
it is clear that as the degree of entanglement
increases, so does ${\bf B_{max}}$ and hence the potential violation.
However without full knowledge of the state being analysed and given that
the two qubit state has a certain degree of entanglement
it is impossible generally to determine the extent to which
the Bell inequality is violated (or for a small degree of
entanglement if it is violated). In terms of
finding other more generalised Bell inequalities that are violated,
the maximally entangled mixed state is a good test candidate. To finish however
the knowledge that the Bell inequality is violated
is strong evidence for the presence of entanglement in the two qubit
system.  The Bell inequality can always be seen as a
{\it marker} for entanglement.

\noindent{\it Acknowledgments}\\
The authors would like to thank D. T. Pope, P. G. Kwiat, N. Gisin and M. Wolf  for useful discussions and
acknowledge the support of the Australian Research Council.    Correspondance can be addressed to Bill Munro,
Hewlett-Packard Laboratories, Filton Road, Stoke Gifford, Bristol BS34 8QZ, U.K.
Electronic address: billm@hplb.hpl.hp.com


\begin{thebibliography}{99}
\bibitem{bellmeasure} If we consider two arbitrary state $\Psi$ and $\psi$,
such that $B(\Psi)>B(\psi)$ the Bell inequality violation for such
states, then it has generally been assumed that the degree of
entanglement for $\Psi$ is greater than $\psi$.
\end{thebibliography}
\end{document}